\documentstyle[12pt,epsfig,rotating]{article}
\DeclareMathAlphabet{\mathsc}{OT1}{cmr}{m}{sc}
\def\21{$SU(2) \otimes U(1) $}
\newcommand{\CL} {C.L.}
\newcommand{\dof}{d.o.f.}

\def\rp{$R_p \hspace{-1em}/\;\:$ }
\def\slash#1{#1\!\!\! /}
\newcommand{\nn}{\nonumber}

\newcommand {\fig} [1] {Fig.~\ref{#1}}

\def\te{{\tilde e}}
\def\tm{{\tilde \mu}}
\def\st{{\tilde \tau}}
 
\def\np#1#2#3{           {Nucl. Phys. }{\bf #1} (19#2) #3}
\def\pl#1#2#3{           {Phys. Lett. }{\bf #1} (19#2) #3}
\def\pr#1#2#3{           {Phys. Rev. }{\bf #1} (19#2) #3}

\begin{document} 
\begin{titlepage} 
\begin{flushright}
hep-ph/0207334 \\ 
IFIC/02-33\\
ZU-TH 11/02 \\ 
\end{flushright} 
\vspace*{3mm} 
\begin{center}  
  \textbf{\large Probing neutrino properties with charged scalar lepton decays}\\[10mm]

{M. Hirsch${}^1$,  W. Porod${}^2$, J. C. Rom\~ao${}^3$ and 
J. W. F. Valle${}^1$ } 
\vspace{0.3cm}\\ 

{\it $^1$ Astroparticle and High Energy Physics Group, 
IFIC - Instituto de F\'\i sica Corpuscular,
Edificio Institutos de Investigaci\'on,
Apartado de Correos 22085,
E-46071 Valencia - Espa\~na \\}
{\it $^2$ Institut f\"ur Theoretische Physik, Universit\"at Z\"urich, \\ 
CH-8057 Z\"urich, Switzerland}

{\it $^3$ Departamento de F\'\i sica, Instituto Superior T\'ecnico\\
          Av. Rovisco Pais 1, $\:\:$ 1049-001 Lisboa, Portugal \\}

\end{center}


\begin{abstract} 
  Supersymmetry with bilinear R-parity violation provides a predictive
  framework for neutrino masses and mixings in agreement with current
  neutrino oscillation data.  The model leads to striking signals at
  future colliders through the R-parity violating decays of the
  lightest supersymmetric particle. Here we study charged scalar
  lepton decays and demonstrate that if the scalar tau is the LSP (i)
  it will decay within the detector, despite the smallness of the
  neutrino masses, (ii) the relative ratio of branching ratios
  $Br({\tilde \tau}_1 \to e \sum \nu_i)/ Br({\tilde \tau}_1 \to \mu
  \sum \nu_i)$ is predicted from the measured solar neutrino angle,
  and (iii) scalar muon and scalar electron decays will allow to test
  the consistency of the model.  Thus, bilinear R-parity breaking SUSY
  will be testable at future colliders also in the case where the LSP
  is not the neutralino.
\end{abstract} 
 
\end{titlepage}

\newpage

\setcounter{page}{1} 

\section{Introduction}

Neutrino physics is one of the most rapidly developing areas of
particle physics \cite{nu2002}. The solar neutrino data, including the
recent measurement of the neutral current rate for solar neutrinos by
the SNO collaboration \cite{Ahmad:2002jz} provide strong evidence for
neutrino flavour conversion. If interpreted in terms of neutrino
oscillations, the data indicate a large mixing angle between $\nu_e$
and $\nu_{\mu}-\nu_{\tau}$, with a strong preference towards the large
mixing angle MSW solution (LMA).  At $3\sigma$ one
has~\cite{Maltoni:2002ni}
\begin{equation}
\label{eq:solbound}
0.25 \le \tan^2\theta_{\odot} \le 0.83  
\end{equation}
for 1 \dof, the best-fit-parameters being 
\begin{equation}
\label{eq:solbfp}
  \tan^2\theta_{\odot} = 0.44 \:\:\:\: 
\Delta m^2_{\odot} = 6.6\times 10^{-5}~{\rm eV^2}
\end{equation}
This nicely confirms earlier hints found in ref.~\cite{Solar}. The 
LMA solution will be testable independently by KamLAND \cite{KamLand},
and first results are expected before the end of the year.  In
addition, current atmospheric neutrino data are most easily explained
by $\nu_{\mu}\leftrightarrow \nu_{\tau}$ oscillations
\cite{Fukuda:1998mi}, with the 3$\sigma$ ranges (1 \dof)
\begin{equation}
\label{eq:atm}
    0.3 \le \sin^2\theta_{Atm} \le 0.7 \,,\quad
    1.2 \times 10^{-3}~{\rm eV^2} \le   \Delta m^2_{\odot}\le 4.8 \times
    10^{-3}~{\rm eV^2} \:.
\end{equation}
These data leave little doubt that neutrinos are massive particles
after all.

Unsurprisingly the discoveries in neutrino oscillation physics have
triggered an avalanche of theoretical and phenomenological papers on
models of neutrino masses and mixings \cite{Avalanche}, the majority
of which are based on one variation or the other of the see-saw
mechanism \cite{seesaw,Mohapatra:1980yp,Schechter:1981cv}. Here we
consider a phenomenologically viable alternative, namely,
supersymmetry with bilinear R-parity breaking terms
\cite{epsrad,e3others}, which in contrast to the seesaw mechanism
generates neutrino masses at the electro-weak scale.  Low-scale
schemes for neutrino masses have the advantage of being potentially
testable in near-future accelerator experiments. In this paper we
study the implications of neutrino physics for charged scalar lepton
decays.

Supersymmetric models with explicit bilinear breaking of R-parity
(BRpV) \cite{epsrad,e3others} provide a simple and calculable
framework for neutrino masses and mixing angles in agreement with the
experimental data \cite{NuMass}. BRpV is a hybrid scheme in which one
neutrino mass is generated at tree-level, through the mixing with the
neutralinos \cite{Ross:1985yg}, in an effective ``low-scale'' variant
of the seesaw, while the remaining two masses are generated at 1-loop
order. A complete 1-loop calculation of the neutrino-neutralino mass
matrix \cite{NuMass} is therefore necessary, before one can confront
the model with experimental data from atmospheric and solar neutrino
experiments. Especially note that the ``solar'' angle has no meaning
in BRpV at tree-level.

BRpV might be considered either as a minimal three-parameter extension
of the MSSM, valid up to some very high energy scale (like the GUT
scale) \cite{expl} or as the effective description of a more
fundamental theory in which the breaking of R-parity is spontaneous
\cite{Ross:1985yg,SBRpV}.  While spontaneous breaking of R-parity may
be considered theoretically more attractive since, for example, it
provides a motivation for the absence of trilinear R-parity breaking
parameters in the superpotential, for the sake of simplicity in our
numerical calculation we will stick to explicit BRpV only.

One should, however, note that the results obtained here are valid
also in those classes of models where R--parity is broken
spontaneously including the presence of an additional Goldstone boson,
namely the Majoron $J$.  This can be seen as follows: The Majoron
consists mainly of the imaginary parts of the \21 singlet scalars,
such as the right-handed sneutrinos \cite{SBRpV}. The only terms which
couple the Majoron directly to sleptons is given by $h^\nu \hat L \hat
H_2 \hat{\nu}^c_R$ in the superpotential and the corresponding term in
the soft SUSY breaking Lagrangian. These terms can in principle induce
decays like $\tilde \tau \to \tilde \mu \, J$.  However, such a decay
requires that one of the charged particles involved contains a large
left-handed component whereas the other one contains a large Higgs
component.  As we will see below, in the cases we will study the
sleptons are mainly right-sleptons. In addition, in mSUGRA scenarios
the mass differences between the lightest three sleptons is rather
small leading to a further suppression of Majoron-emitting charged
slepton decays.

If R-parity is broken the lightest supersymmetric particle (LSP) will
decay. As was shown in \cite{NtrlDecay} (see also
\cite{acceltestnew}), if the LSP is the lightest neutralino, the
measured low-energy neutrino properties translate into predictions for
the ratios of various branching ratios of the neutralino decay, thus
providing a definite test of the model as the origin of neutrino
masses and mixings.

However, cosmological and astrophysical constraints on its nature no
longer apply if the LSP decays. Thus, within R-parity violating SUSY a
priori {\em any} superparticle could be the LSP. In this paper we
study the case where a charged scalar lepton, most probably the scalar
tau, is the LSP~\footnote{The case of light stop decays was considered
  in ref.~\cite{Restrepo:2001me,Allanach:1999bf}}. We calculate the
production and decays of $\st$, as well as the decays of $\te$ and
$\tm$, and demonstrate that also for the case of charged sleptons as
LSPs neutrino physics leads to definite predictions of various decay
properties.

This paper is organized as follows. In Sec. \ref{sec:model} we will
define the model, discuss the charged scalar mass matrix and give some
formulas for the two-body decays of charged sleptons, which are the
most important decay channels. In Sec. \ref{sec:slept-prod-decays} we
will then discuss production and decays of these particles, with
special emphasis on possible measurements of R-parity violating
parameters. Finally, in Sect. \ref{sec:conclusions} we summarize our
conclusions.

\section{The model}
\label{sec:model}

Since BRpV SUSY has been discussed in the literature several times
\cite{epsrad,e3others,NuMass,BilBB1} we will repeat only the main features 
of the model here.  We will follow the notation of \cite{NuMass}.

The simplest bilinear \rp model (we call it the \rp MSSM) is
characterized by three additional terms in the superpotential
\begin{equation}
\label{eq:Wpot} 
W = W_{MSSM} + W_{\slash R_P} 
\end{equation} 
where $W_{MSSM}$ is the ordinary superpotential of the MSSM and
\begin{equation}
\label{eq:WRPV} 
W_{\slash R_P} = \epsilon_i \widehat
  L_i\widehat H_u.  
\end{equation} 
These bilinear terms, together with the corresponding terms in the
soft SUSY breaking part of the Lagrangian, 
\begin{equation}
\label{eq:Lsoft}
  {\cal L}_{soft} = {\cal L}_{soft}^{MSSM} + B_i \epsilon_i {\tilde
    L}_i H_u 
\end{equation} 
define the minimal model, which we will adopt throughout this paper.
The appearance of the lepton number violating terms in Eq.
(\ref{eq:Lsoft}) leads in general to non-zero vacuum expectation
values for the scalar neutrinos $\langle {\tilde \nu}_i \rangle$,
called $v_i$ in the rest of this paper, in addition to the VEVs $v_U$
and $v_D$ of the MSSM Higgs fields $H_u^0$ and $H_d^0$.  Together with
the bilinear parameters $\epsilon_i$ the $v_i$ induce mixing between
various particles which in the MSSM are distinguished (only) by lepton
number (or R--parity).  Mixing between the neutrinos and the
neutralinos of the MSSM, as mentioned previously, generates a non-zero
mass for one specific linear superposition of the three neutrino
flavour states of the model at tree-level. For a complete discussion of 
1-loop corrections, providing mass for the remaining two neutrino states, 
see \cite{NuMass}.

For the decays of the charged sleptons it is necessary to calculate the 
mixings between neutrinos and neutralinos, charginos and charged 
leptons, as well as the charged scalar mixing. Since the various 
mass matrices can be found in \cite{NuMass}, we will discuss only the 
charged scalar mass matrix in the next section. 

\subsection{The charged scalar mass matrix}
\label{sec:charged-scalar-mass}

With R-parity broken by the bilinear terms in Eq. (\ref{eq:WRPV}) the
left-handed and right-handed charged sleptons mix with the charged
Higgs of the MSSM, resulting in an ($8 \times 8$) mass matrix for
charged scalars. As in the MSSM this matrix contains the Goldstone
boson, providing the mass of the W-boson after electro-weak symmetry
breaking. One can rotate away the Goldstone mode from this mass
matrix, using the following rotation matrix

\begin{equation}
\label{eq:Rhat}
{\hat R} = 
\left[
\begin{array}{cccccccc}
\frac{v_D}{w_3} & -\frac{v_U}{w_3} & \frac{v_1}{w_3} & 
\frac{v_2}{w_3} &\frac{v_3}{w_3} & 0 & 0 & 0\cr
\frac{v_U}{w_0} & \frac{v_D}{w_0} & 0 & 
0 & 0 & 0 & 0 & 0\cr
- \frac{v_1 v_D}{w_0 w_1} & \frac{v_1 v_U}{w_0 w_1} & \frac{w_0}{w_1} & 
0 & 0 & 0 & 0 & 0\cr
- \frac{v_2 v_D}{w_1 w_2} & \frac{v_2 v_U}{w_1 w_2} & 
-\frac{v_2 v_1}{w_1 w_2} & \frac{w_1}{w_2} & 0 & 0 & 0 & 0\cr
- \frac{v_3 v_D}{w_2 w_3} & \frac{v_3 v_U}{w_2 w_3} & 
-\frac{v_3 v_1}{w_2 w_3} & -\frac{v_2 v_3}{w_2 w_3} & 
\frac{w_2}{w_3} & 0 & 0 & 0 \cr
0 & 0 & 0 & 0 & 0 & 1 & 0 & 0 \cr
0 & 0 & 0 & 0 & 0 & 0 & 1 & 0 \cr
0 & 0 & 0 & 0 & 0 & 0 & 0 & 1 \cr
\end{array}
\right]
\end{equation}
where,

\begin{eqnarray}
  \label{eq:shorthands}
  w_0 = \sqrt{v_D^2+v_U^2} \\
w_1 = \sqrt{v_1^2+v_D^2+v_U^2} \\
w_2 = \sqrt{v_1^2+v_2^2+v_D^2+v_U^2} \\
w_3 = \sqrt{v_1^2+v_2^2+v_3^2+v_D^2+v_U^2} 
\end{eqnarray}

This matrix has the property that
\begin{eqnarray}
  \label{eq:rot}
{\hat R} M_{S^{\pm}}^2 {\hat R}^T = \left[  
\begin{array}{cc}  
0 & {\vec 0}^T \cr
{\vec 0} & M_{S_7^{\pm}}^2 \cr
\end{array}
\right]
\end{eqnarray}
where $M_{S_7^{\pm}}^2$ is a ($7\times 7$) matrix and the zeroes  
in the first row and first column correspond to the (massless) Goldstone 
state in $\xi = 0$ gauge.

We divide the remaining $M_{S_7^{\pm}}^2$ into two parts,
\begin{equation}
\label{eq:splitMSC}
M_{S_7^{\pm}}^2 = (M_{S_7^{\pm}}^2)^{(0)} +(M_{S_7^{\pm}}^2)^{(1)}
\end{equation}
where $(M_{S_7^{\pm}}^2)^{(0)}$ [$(M_{S_7^{\pm}}^2)^{(1)}$] contains 
only R-parity conserving [R-parity violating] terms. 
Note that in the following we assume for simplicity that there is no 
inter-generational mixing among the charged sleptons. This is motivated 
by existing constraints from flavour changing neutral currents \cite{FCNC} 
and is consistent with the minimal SUGRA scenario of the MSSM, which we 
will use in the numerical part of this paper. With this assumption  
also the branching ratio $\mu \rightarrow e \gamma$ is small \cite{MuEGam} 
in the bilinear model in agreement with experimental data.
The R-parity conserving part of $M_{S_7^{\pm}}^2$ is given by
\begin{equation}
\label{eq:ChSc0}
(M_{S_7^{\pm}}^2)^{(0)}=
\left[
\begin{array}{ccccccc}
m_{H^{\pm}}^2 & \cdot &\cdot &\cdot &\cdot &\cdot &\cdot \cr
0 & {\hat m}_{L_1}^2 & \cdot & \cdot& \cdot& \cdot  & \cdot \cr
0 & 0 & {\hat m}_{L_2}^2 & \cdot& \cdot& \cdot  & \cdot \cr
0 & 0 & 0 & {\hat m}_{L_3}^2 & \cdot& \cdot  & \cdot \cr
0 & {\hat m}_{LR1}^2 & 0 & 0 & {\hat m}_{R_1}^2 & \cdot  & \cdot \cr
0 & 0 & {\hat m}_{LR2}^2 & 0 & 0 & {\hat m}_{R_2}^2   & \cdot \cr
0 & 0 & 0 & {\hat m}_{LR3}^2 & 0 & 0 & {\hat m}_{R_3}^2  \cr
\end{array}
\right]
\end{equation}
where the dots indicate that the matrix is symmetric and 
\begin{equation}
\label{eq:defhmpm}
m_{H^{\pm}}^2 = m_A^2 + \frac{g^2 v_{R_P}^2}{4}
\end{equation}
\begin{equation}
\label{eq:defhmli}
{\hat m}_{L_i}^2 = m_{L_i}^2 - (g^2 - {g'}^2)\frac{v_{R_P}^2}{8}c_{2\beta} 
                 + \frac{1}{2}(h_{i}^E)^2 v_D^2
\end{equation}
\begin{equation}
\label{eq:defhmri}
{\hat m}_{R_i}^2 = m_{R_i}^2 - {g'}^2\frac{v_{R_P}^2}{4}c_{2\beta} 
                 + \frac{1}{2}(h_{i}^E)^2 v_D^2
\end{equation}
\begin{equation}
\label{eq:defhmlri}
{\hat m}_{LRi}^2 = + \frac{1}{\sqrt{2}}(h_{i}^E)(A_i v_D - \mu v_U)
\end{equation}
with $v_{R_P}^2=v_U^2+v_D^2$. $m_A^2$ is the MSSM pseudoscalar Higgs
mass parameter $m_A^2 = (\mu B)/(s_{\beta}c_{\beta})$, $h_i^E$ and
$A_i$ are the Yukawa couplings and soft breaking trilinear parameters
of the charged lepton of generation $i$, $\mu$ is the Higgsino mixing
parameter characterizing the superpotential, and $c_{2\beta} =
\cos(2\beta)$, where $\beta$ is defined in the usual way as $\tan\beta
= v_U/v_D$.  The R-parity violating part of $M_{S_7^{\pm}}^2$
can be written as

\begin{equation}
\label{eq:ChSc1}
(M_{S_7^{\pm}}^2)^{(1)} =
\left[
\begin{array}{ccc}
\Delta m_{H^{\pm}}^2 & ({\vec X}_{HL})^T & ({\vec X}_{HR})^T \cr
{\vec X}_{HL} & M_{LL}^{2(1)} & (M_{LR}^{2(1)})^T \cr
{\vec X}_{HR} & M_{LR}^{2(1)} & M_{RR}^{2(1)} 
\end{array} 
\right] .
\end{equation}
The Higgs mass correction and the Higgs-Slepton mixing terms in eq.
(\ref{eq:ChSc1}) are

\begin{equation}
\label{eq:DeltamHp}
\Delta m_{H^{\pm}}^2 = \sum {\Big\{ }
                   (\frac{v_i}{v_D})^2 {\bar m}_{\tilde \nu_i}^2
                        \frac{c_{\beta}^4}{s_{\beta}^2}
                     -  \epsilon_i \mu \frac{v_i}{v_D} 
                        \frac{c_{2\beta}}{s_{\beta}^2}
                     + \frac{g^2}{4} v_i^2 c_{2\beta}
                     + \frac{1}{2}  (h_i^E v_i)^2 s_{\beta}^2{\Big\} }
\end{equation}
\begin{equation}
\label{eq:DelMHLi}
(X_{HL})_i = \frac{v_i}{v_D} {\bar m}_{\tilde \nu_i}^2
                        \frac{c_{\beta}^2}{s_{\beta}}
                     - \mu \epsilon_i \frac{1}{s_{\beta}}
             +\frac{1}{2} (g^2 - (h_i^E)^2) v_D v_i s_{\beta}
\end{equation}
\begin{equation}
\label{eq:DelMHRi}
(X_{HR})_i = - \frac{1}{\sqrt{2}}h_i^E v_i (A_is_{\beta} + \mu c_{\beta})
             - \frac{1}{\sqrt{2}}h_i^E \epsilon_i v_D \frac{1}{c_{\beta}}
\end{equation}

$M_{LL}^{2(1)}$ can be written as,
\begin{equation}\label{DefMLL21}
M_{LL}^{2(1)}= 
\left[
\begin{array}{ccc}
\Delta m_{L_1}^2 & (X_{LL})_{12} & (X_{LL})_{13} \cr
(X_{LL})_{12} & \Delta m_{L_2}^2 & (X_{LL})_{23} \cr
(X_{LL})_{13} & (X_{LL})_{23} &  \Delta m_{L_3}^2
\end{array}
\right]
\end{equation}
with the diagonal terms given by
\begin{equation}
\label{eq:DefMLLii}
\Delta m_{L_i}^2 = (\frac{v_i}{v_D})^2 {\bar m}_{\tilde \nu_i}^2 c_{\beta}^2 +
 \epsilon_i^2 + \frac{1}{2}  (g^2 + (h_i^E)^2) v_i^2 c_{\beta}^2
                  + \frac{1}{8} (g^{\prime 2} - g^2) \sum v_i^2
\end{equation}
whereas the off-diagonals are
\begin{eqnarray}
  \label{eq:xllij}
(X_{LL})_{12} = \epsilon_1 \epsilon_2 
        &+& (\frac{v_1}{v_D}) (\frac{v_2}{v_D})m_{L_2}^2 c_{\beta}^2 \\ \nn
        &+& v_1 v_2 \Big[\frac{1}{4}(g^2 + (h_2^E)^2) -
                          \frac{1}{8} (g^2 -  g^{\prime 2}) c_{2\beta} +
                          \frac{1}{4}(h_2^E)^2 c_{2\beta}\Big] \\
(X_{LL})_{13} = \epsilon_1 \epsilon_3 
        &+& (\frac{v_1}{v_D}) (\frac{v_3}{v_D})m_{L_3}^2 c_{\beta}^2 \\ \nn
        &+& v_1 v_3 \Big[\frac{1}{4}(g^2 + (h_3^E)^2) -
                          \frac{1}{8} (g^2 -  g^{\prime 2}) c_{2\beta} +
                          \frac{1}{4}(h_3^E)^2 c_{2\beta}\Big] \\
(X_{LL})_{23} = \epsilon_2 \epsilon_3 
        &+& (\frac{v_2}{v_D}) (\frac{v_3}{v_D})m_{L_3}^2 c_{\beta}^2 \\ \nn
        &+& v_2 v_3 \Big[\frac{1}{4}(g^2 + (h_3^E)^2) -
                          \frac{1}{8} (g^2 -  g^{\prime 2}) c_{2\beta} +
                          \frac{1}{4}(h_3^E)^2 c_{2\beta}\Big]
\end{eqnarray}
Similarly for $M_{RR}^{2(1)}$, 
\begin{equation}
\label{eq:defmrrii}
\Delta m_{R_i}^2 = \frac{1}{2} (h_i^E)^2 v_i^2 -
                    \frac{1}{4}  g^{\prime 2} \sum v_i^2
\end{equation}
and
\begin{equation}\label{defmrrij}
(X_{RR})_{ij}=\frac{1}{2} (h_i^E)(h_j^E) v_i v_j
\end{equation}
Finally, the matrix
$M_{LR}^{2(1)}$ has the following peculiar structure,

\begin{equation}\label{DefMLR21}
M_{LR}^{2(1)}= 
\left[
\begin{array}{ccc}
(X_{LR})_{11} & 0 & 0 \cr
(X_{LR})_{12} & (X_{LR})_{22} & 0 \cr
(X_{LR})_{13} & (X_{LR})_{23} & (X_{LR})_{33}
\end{array}
\right]
\end{equation}
where
\begin{equation}\label{DefXLRii}
(X_{LR})_{ii} = - \frac{1}{2\sqrt{2}}(h_i^E)(\frac{v_i}{v_D})^2 c_{\beta} 
                  v_D \Big[ \mu s_{\beta} - A_i c_{\beta} \Big]
\end{equation}
\begin{equation}\label{DefXLRij}
(X_{LR})_{ij} = - \frac{1}{\sqrt{2}}(h_i^E)(\frac{v_i}{v_D})
                   (\frac{v_j}{v_D}) c_{\beta} v_D 
                  \Big[ \mu s_{\beta} - A_i c_{\beta} \Big]
\end{equation}
In the above equations we have used the following abbreviation
\begin{equation}\label{defmsnu}
{\bar m}_{\tilde \nu_i}^2 = m_{L_i}^2 + \frac{1}{8} (g^2 +  g^{\prime 2})
                                                    (v_D^2-v_U^2).
\end{equation}
With the definitions outlined above, once can easily derive approximate 
expressions for the mixing between the charged Higgs and the charged 
sleptons induced by the R-parity breaking parameters. These are given 
by
\begin{equation}\label{Lmix}
\sin\theta_{HL_i} \simeq \frac{X_{HL,i}}{(m^2_{H^{\pm}}-m^2_{L_i})},
\end{equation}
\begin{equation}\label{Rmix}
\sin\theta_{HR_i} \simeq \frac{X_{HR,i}}{(m^2_{H^{\pm}}-m^2_{R_i})}.
\end{equation}
Note that one expects $\sin\theta_{HR_i} \sim h^E_i
\sin\theta_{HL_i}$, i.e. the mixing between right-handed sleptons and
the Higgs should be typically much smaller than the left-handed
Higgs-slepton mixing.

Finally, the R-parity conserving mixing between left-handed and 
right-handed sleptons is approximately given by 

\begin{equation}\label{LRmix}
\sin 2\theta_{\tilde l_i} \simeq \frac{2 {\hat m}_{LRi}^2}
                       {{\hat m}_{Li}^2-{\hat m}_{Ri}^2}.
\end{equation}

\subsection{Formulas for two-body decays}
\label{sec:formulas-two-body}

Charged scalar leptons lighter than all other supersymmetric particles
will decay through R-parity violating couplings. Possible final states
are either $l_j\nu_k$ or $q{\bar q}'$. For right-handed charged
sleptons (${\tilde l}_{Ri}$) the former by far dominates over the
hadronic decay mode, since the mixing between ${\tilde l}_{Ri}$ and
the charged Higgs is small, as explained above.

In the limit $(m_{f_j},m_{\nu_k}) \ll m_{{\tilde f}_i}$ one has the
simple formula for the two-body decays ${\tilde f}_i \rightarrow f_j +
\nu_k$,
\begin{equation}\label{width}
\Gamma_{{\tilde f}_if_j\nu_k} = \frac{m_{{\tilde f}_i}}{16\pi} 
\Big[(O^{cns}_{Lf_j\nu_k{\tilde f}_i})^2 + 
(O^{cns}_{Rf_j\nu_k{\tilde f}_i})^2\Big]
\end{equation}
Exact expressions for these couplings can be found, for
example, in ref.~\cite{NuMass}. Even though in the results presented
in this paper we have always calculated the couplings appearing in
eq. (\ref{width}) exactly using our numerical code, it is instructive to
consider an approximate diagonalization procedure for the various mass
matrices.  This method is based on the fact that
neutrino masses are much smaller than all other particle masses in the
theory and therefore one expects that the bilinear R-parity breaking
parameters are (somewhat) smaller than the corresponding MSSM
parameters. For the charged scalar mass matrix all necessary
definitions have been given above, for details for the corresponding
procedure for neutralino and chargino mass matrices we refer to
\cite{NuMass,BilBB1,Now96}.

For the case where $i \ne j$ for ${\tilde l}_{Ri} \rightarrow l_j \sum
\nu_k$ one finds
\begin{eqnarray}
\label{eq:SlLNu}
\sum_k \Big[(O^{cns}_{L l_j\nu_k{\tilde l}_{i}})^2 + 
(O^{cns}_{Rl_j \nu_k {\tilde l}_{i}})^2 \Big] &=& 
(-h^E_{l_i}c_{\tilde l_i}\frac{\epsilon_j}{\mu}- (gs_{\tilde l_i}y_1+
h^E_{l_i}c_{\tilde l_i} y_2) \Lambda_{j})^2 \\ \nn
& + & (h^E_{l_j})^2 (s_{\beta} \sin\theta_{HR_i}-c_{\beta}^2 
 s_{\tilde l_i} {\tilde v}_i)^2   
\end{eqnarray}
\begin{eqnarray}
\label{eq:SimSlLNu}
&\simeq & 
(c_{\tilde l_i}h^E_{l_i}\frac{\epsilon_j}{\mu})^2
\end{eqnarray}
Here $c_{\tilde l_i} \equiv \cos(\theta_{\tilde l_i})$ and $s_{\tilde
  l_i} \equiv \sin(\theta_{\tilde l_i})$ where $\theta_{\tilde l_i}$
is the left-right mixing angle for ${\tilde l_i}$, $\sin\theta_{HR_i}$
characterizes the charged Higgs-(right-handed)-Slepton mixing and
${\vec \Lambda}$ is given by
\begin{equation}
\label{deflam}
\Lambda_i = \epsilon_i v_D + \mu v_i.
\end{equation}
The quantities $y_1$ and $y_2$ are defined as

\begin{eqnarray}\label{eq:defy}
y_1 = \frac{g}{\sqrt{2}{\rm Det}M_{\chi^{\pm}}} \\
y_2 =- \frac{g^2 v_U}{2\mu{\rm Det}M_{\chi^{\pm}}}
\end{eqnarray}
with ${\rm Det}M_{\chi^{\pm}}$ being the determinant of the MSSM 
chargino mass matrix.

While eq. (\ref{eq:SlLNu}) above keeps all R-parity breaking paramters in
the expansion up to second order, eq. (\ref{eq:SimSlLNu}) should be valid
in the parameter region in which the 1-loop neutrino masses are
smaller than the tree-level contribution.

For the case $i=j$ the corresponding formulas are rather cumbersome
and therefore of limited utility, except for the case ${\tilde l} =
\te$.  Here, since $h_e \ll 1$ one can simplify the couplings to,

\begin{equation}\label{SeENu}
\sum_k \Big[(O^{cns}_{Le{\nu_k}{\tilde e}})^2 + 
(O^{cns}_{Re{\nu_k}{\tilde e}})^2 \Big] 
\simeq 2  g^{\prime 2} x_1^2 |{\vec \Lambda}|^2 
\end{equation}
The parameter ${\vec \Lambda}$ has been defined above and $x_1$ is
given by
\begin{equation}
\label{defx1}
x_1 = \frac{g' M_2 \mu}{2 {\rm Det}M_{\chi^0}}
\end{equation}
with ${\rm Det}M_{\chi^0}$ being the determinant of the MSSM neutralino 
mass matrix and $M_2$ the soft SUSY breaking $SU(2)$ mass parameter.

From eq. (\ref{eq:SimSlLNu}) one expects that various ratios of branching
ratios should contain rather precise information on ratios of the
bilinear R-parity breaking parameters, for example, $Br({\tilde
  \tau}_1 \rightarrow e \sum \nu_i)/ Br({\tilde \tau}_1 \rightarrow
\mu \sum \nu_i) \simeq (\epsilon_1/\epsilon_2)^2$. We will discuss
this important point in more detail in the next section.

\section{Slepton production and decays}
\label{sec:slept-prod-decays}

In this section we will discuss charged slepton production and decay
modes. In order to reduce the number of parameters, the numerical
calculations were performed in the mSUGRA version of the MSSM. Unless
noted otherwise, we have scanned the parameters in the following
ranges: $M_2$ from [0,1.2] TeV, $|\mu|$ from [0,2.5] TeV, $m_0$ in the
range [0,0.5] TeV, $A_0/m_0$ and $B_0/m_0$ [-3,3] and $\tan\beta$
[2.5,10].  All randomly generated points were subsequently tested for
consistency with the minimization (tadpole) conditions of the Higgs
potential as well as for phenomenological constraints from
supersymmetric particle searches. In addition, we selected points in
which at least one of the charged sleptons was lighter than the
lightest neutralino, and thus the LSP. This latter cut prefers
strongly $m_0 <<M_2$.

R-parity violating parameters were chosen in such a way \cite{NuMass}
that the neutrino masses and mixing angles are approximately
consistent with the experimental data. A good ``fit'' to the data
would require: a) $\Lambda_{\mu} \simeq \Lambda_{\tau}$, in order to
account for a nearly maximal $\nu_{\mu} \rightarrow \nu_{\tau}$
angle in atmospheric oscillations,~eq.~\ref{eq:atm}; b) $\Lambda_{e}
< \Lambda_{\tau}$, to fulfil the constraints from $\nu_e$-oscillation
searches at reactors~\cite{Apollonio:1999ae}; c) $|\vec \Lambda|
\simeq [0.05,2]$ GeV$^2$, for the atmospheric neutrino mass
scale,~eq.~\ref{eq:atm}; d) $\epsilon_1 \simeq \epsilon_2$, to have
a large angle in solar oscillations,~eq.~\ref{eq:solbfp}; and e)
$|{\vec \epsilon}|^2/|\vec \Lambda| \simeq [0.1,10]$, for the solar
mass scale,~eq.~\ref{eq:solbfp}.

In order to investigate the dependence of our results on the
assumptions about the R-parity violating parameters, we construct
three different sets of points. Set1 was calculated to give an
approximate ``fit'' to the neutrino data, as described above. Set2 is
similar to Set1, except that $\epsilon_1/\epsilon_2$ has been varied
in a wider range ([0.1,10]), so as to cover both large and small solar
angles~\footnote{Although at the moment the small angle solar solution
  is ruled out by a careful analysis of the solar
  data~\cite{Maltoni:2002ni}, it does not cost us much additional
  effort to keep this option in mind.}. 
The last set, called Set3 in the following, is again similar to Set1,
except that $\epsilon_2/\epsilon_3$, which is hardly constrained by
neutrino data, is varied in the interval $\epsilon_2/\epsilon_3 \simeq
0.1-2$.

In supersymmetric models in which the scalar leptons have a common
soft SUSY breaking mass parameter at some high scale ($m_0$ in mSUGRA)
the renormalization group evolution leads to some splitting between the
scalar taus and the $\te$ and $\tm$ states at the weak scale. While
the lightest mass eigenstate in the charged slepton sector is usually
mainly the $\st_R$, the eigenvalues for $\te_R$ and $\tm_R$ are not
much heavier, such that also $\te_R$ and $\tm_R$ decay mainly via
R-parity violating two-body decays. In our numerical calculation we
therefore not only consider the decays of $\st_R$, but also those of
$\te_R$ and $\tm_R$. These decays can provide information on the
R-parity violating parameters not accesible in $\st_R$ decays and
allow for additional cross checks of the consistency of the model.
This is true especially for the case of lepton flavor violating
slepton decays since from eq.~\ref{eq:SimSlLNu} one expects them to
be directly correlated with the BRpV parameters $\epsilon_i$.
\begin{figure}
\setlength{\unitlength}{1mm}
\begin{center}
\begin{picture}(80,70)
\put(0,0){\mbox{\epsfig{figure=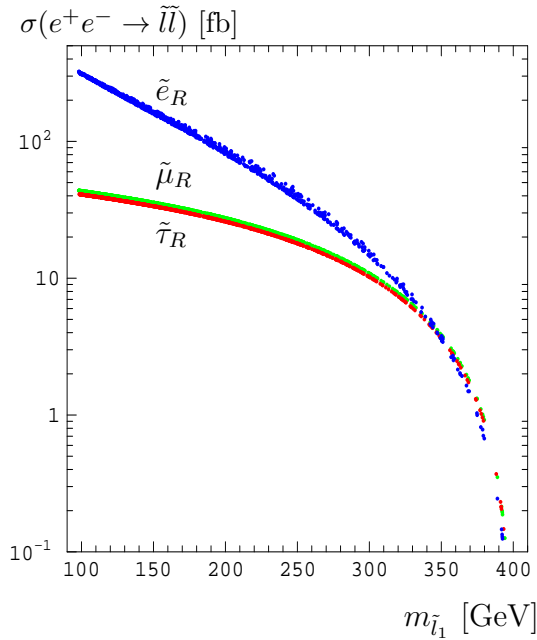,height=7.7cm,width=7.cm}}}
\put(2,76){\makebox(0,0)[bl]
            {{\small $\sigma(e^+ e^- \to {\tilde l}{\tilde l})$~[fb]}}}
\put(71,-3){\makebox(0,0)[br]{{$m_{\tilde l_1}$ [GeV]}}}
\put(20,67){\mbox{${\tilde e}_R$}}
\put(20,56){\mbox{${\tilde \mu}_R$}}
\put(20,48){\mbox{${\tilde \tau}_R$}}
\end{picture}
\end{center}
\caption[]{$e^+ e^- \to {\tilde l}{\tilde l}$ production 
  cross section as a function of $m_{\tilde l}$ at a Linear Collider
  with 0.8 TeV c.m.s energy.  From top to bottom: ${\tilde e}$ (dark,
  on colour printers blue), ${\tilde \mu}$ (light shaded, green) and
  ${\tilde \tau}$ (dark shaded, red).}
\label{fig:Prod0.8TeV}
\end{figure}

\begin{figure}
\setlength{\unitlength}{1mm}
\begin{center}
\begin{picture}(80,50)
\put(0,-30){\mbox{\epsfig{figure=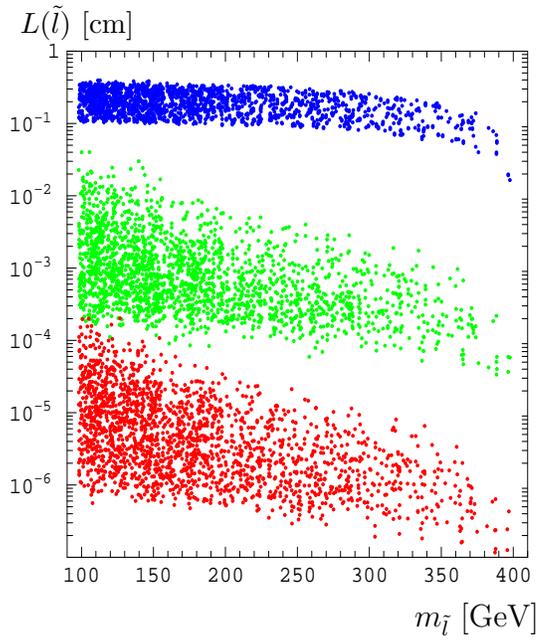,height=7.7cm,width=7.cm}}}
\put(2,46){\makebox(0,0)[bl]
            {{\small $L({\tilde l})$~[cm]}}}
\put(71,-33){\makebox(0,0)[br]{{$m_{\tilde l}$ [GeV]}}}
\end{picture}
\end{center}
\vskip28mm
\caption[]{Charged slepton decay length 
  as a function of $m_{\tilde l}$ at a linear collider with 0.8 TeV
  c.m.s.  energy. From top to bottom: ${\tilde e}$ (dark, on color
  printers blue), ${\tilde \mu}$ (light shaded, green) and ${\tilde
    \tau}$ (dark shaded, red).}
\label{fig:DecayLength}
\end{figure}

For the calculation of the cross section we have adapted the
formulas given in \cite{Blochinger:2002zw} to the bilinear model
taking into account correctly all mixing effects in the numerical
calculation. 
In \fig{fig:Prod0.8TeV} we show the cross section $\sigma(e^+ e^- \to
{\tilde l}{\tilde l})$ in [fb] for $\sqrt{s}=0.8$ TeV as a
function of the charged scalar mass, for ${\tilde e}$, ${\tilde \mu}$
and ${\tilde \tau}$, respectively. Assuming an integrated luminosity
of 1000 fb$^{-1}$ per year can be achieved at a future linear collider
\cite{tesla,Aguilar-Saavedra:2001rg} this implies that around $10^4$ 
scalar muons and 
scalar taus can be directly produced per year. For scalar electrons
one expects between $10^4$ and $10^5$ produced pairs per year. Since
the three R-parity violating two-body decay channels of the
right-handed sleptons nearly add up to 100 \%, one can expect that
individual branching ratios will be measured to an accuracy of 1 \% if
they occur with similar strength.

At the LHC the direct production of right--sleptons is small. As a
result, they will be produced mainly in cascade decays. The relative
$\te_R$, $\tm_R$ and $\st_R$ yields will depend on the details of the
cascade decays involved. Let us consider for simplicity the case where
the cascade decays of the coloured particles end up in the lightest
neutralino as in the MSSM.  Beside the kinematics, the resulting
number of $\te_R$, $\tm_R$ and $\st_R$ arising from these decays
depends on the nature of the lightest neutralino. When this is mainly
bino-like, one expects that it decays dominantly into an equal number
of $\te_R$, $\tm_R$ and $\st_R$'s. As a result the number of
right-sleptons roughly equal to the number of neutralinos. Also in
case of a wino--like neutralino the amount of $\te_R$, $\tm_R$ and
$\st_R$ will be equal. However, in this case the main lightest
neutralino decay mode will be to a $W$--boson and a charged lepton,
leaving fewer sleptons to be studied. However, as discussed in
\cite{NtrlDecay,acceltestnew}, in this case the neutralino decay modes
can be used to probe the large atmospheric neutrino angle. For the case
where the lightest neutralino is higgsino--like it will decay into a
$W$--boson and a charged lepton, or into a $Z$--boson and a neutrino,
similar to the wino case. However for large $\tan\beta$ the decay into
$\st_R$ will again be important, even for higgsino--like neutralinos.

In \fig{fig:DecayLength} we show the charged scalar leptons decay
length ($\te$, $\tm$ and $\st$, from top to bottom) as a function of
the scalar lepton masses for Set3. Very similar results hold for the
other sets which are therefore not shown. All decay lengths are small
compared to typical detector sizes, despite the smallness of the
neutrino masses.  The three generations of sleptons decay with quite
different decay lengths and thus it should be possible to separate the
different generations experimentally at a future linear collider.
Note that the ratio of the decay lengths $L(\st)/L(\tm)$ is
approximately given by $(h_{\mu}/h_{\tau})^2$.

\begin{figure}
\vskip-30mm
\hskip5mm
\epsfysize=140mm
\epsfxsize=100mm
\epsfbox{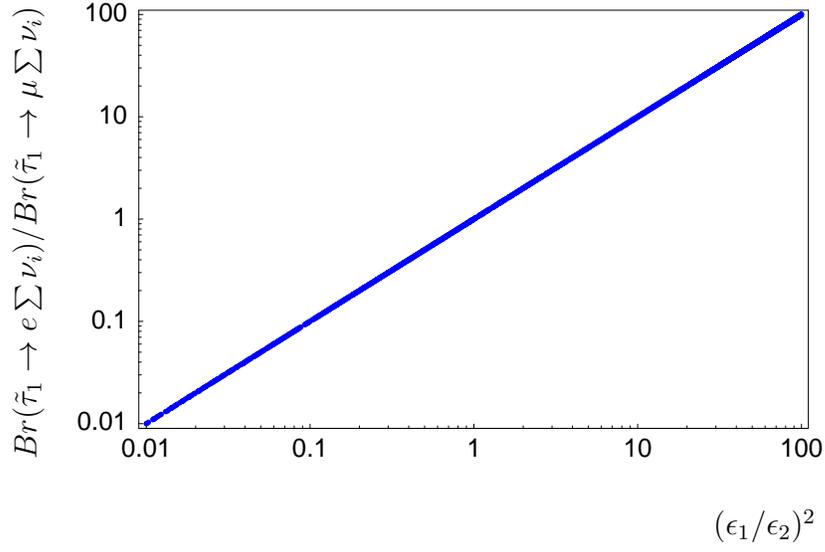}

\vskip-45mm
\begin{rotate}{90}
$Br({\tilde \tau}_1 \rightarrow e \sum \nu_i)/
Br({\tilde \tau}_1 \rightarrow \mu \sum \nu_i)$
\end{rotate}

\vskip5mm
\hskip90mm
$(\epsilon_1/\epsilon_2)^2$

\vskip-30mm
\hskip5mm
\epsfysize=140mm
\epsfxsize=100mm
\epsfbox{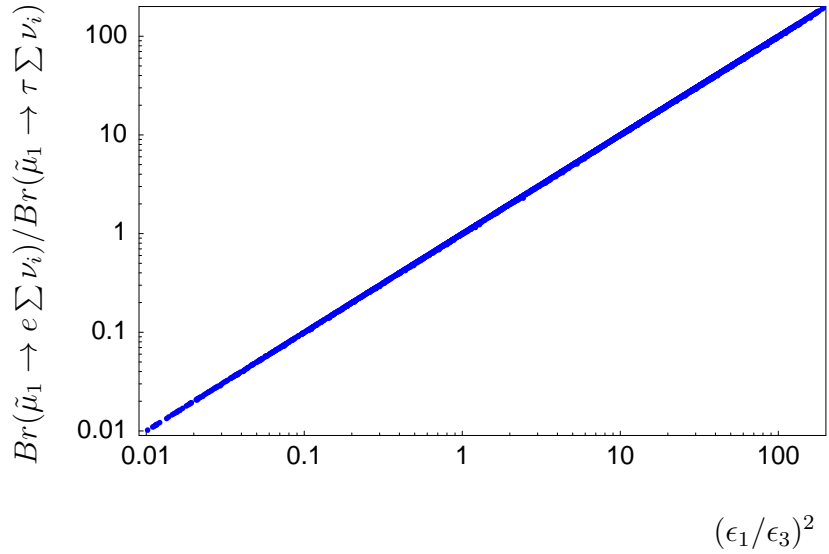}

\vskip-45mm
\begin{rotate}{90}
$Br({\tilde \mu}_1 \rightarrow e \sum \nu_i)/
Br({\tilde \mu}_1 \rightarrow \tau \sum \nu_i)$
\end{rotate}

\vskip5mm
\hskip90mm
$(\epsilon_1/\epsilon_3)^2$

\caption[]{Ratios of branching ratios for scalar tau  decays (top panel)
  versus $(\epsilon_1/\epsilon_2)^2$, and scalar muon decays (bottom
  panel) versus $(\epsilon_1/\epsilon_3)^2$ for Set2.}
\label{BrSet4a}
\end{figure}

\begin{figure}
\vskip-30mm
\hskip5mm
\epsfysize=140mm
\epsfxsize=100mm
\epsfbox{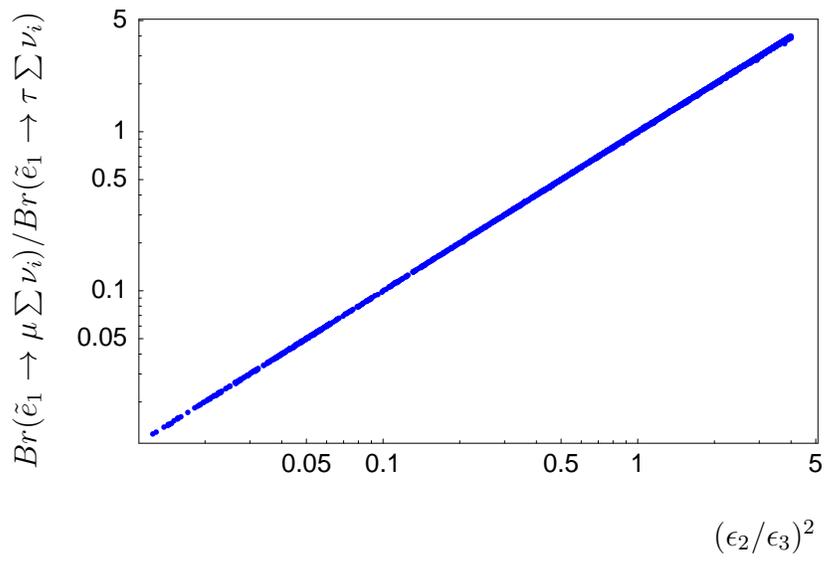}

\vskip-45mm
\begin{rotate}{90}
$Br({\tilde e}_1 \rightarrow \mu \sum \nu_i)/
Br({\tilde e}_1 \rightarrow \tau \sum \nu_i)$
\end{rotate}

\vskip5mm
\hskip90mm
$(\epsilon_2/\epsilon_3)^2$

\caption[]{Ratios of branching ratios for scalar electron  decays 
versus $(\epsilon_2/\epsilon_3)^2$ for Set3.}
\label{BreSet4b}
\end{figure}

\begin{figure}
\vskip-30mm
\hskip5mm
\epsfysize=140mm
\epsfxsize=100mm
\epsfbox{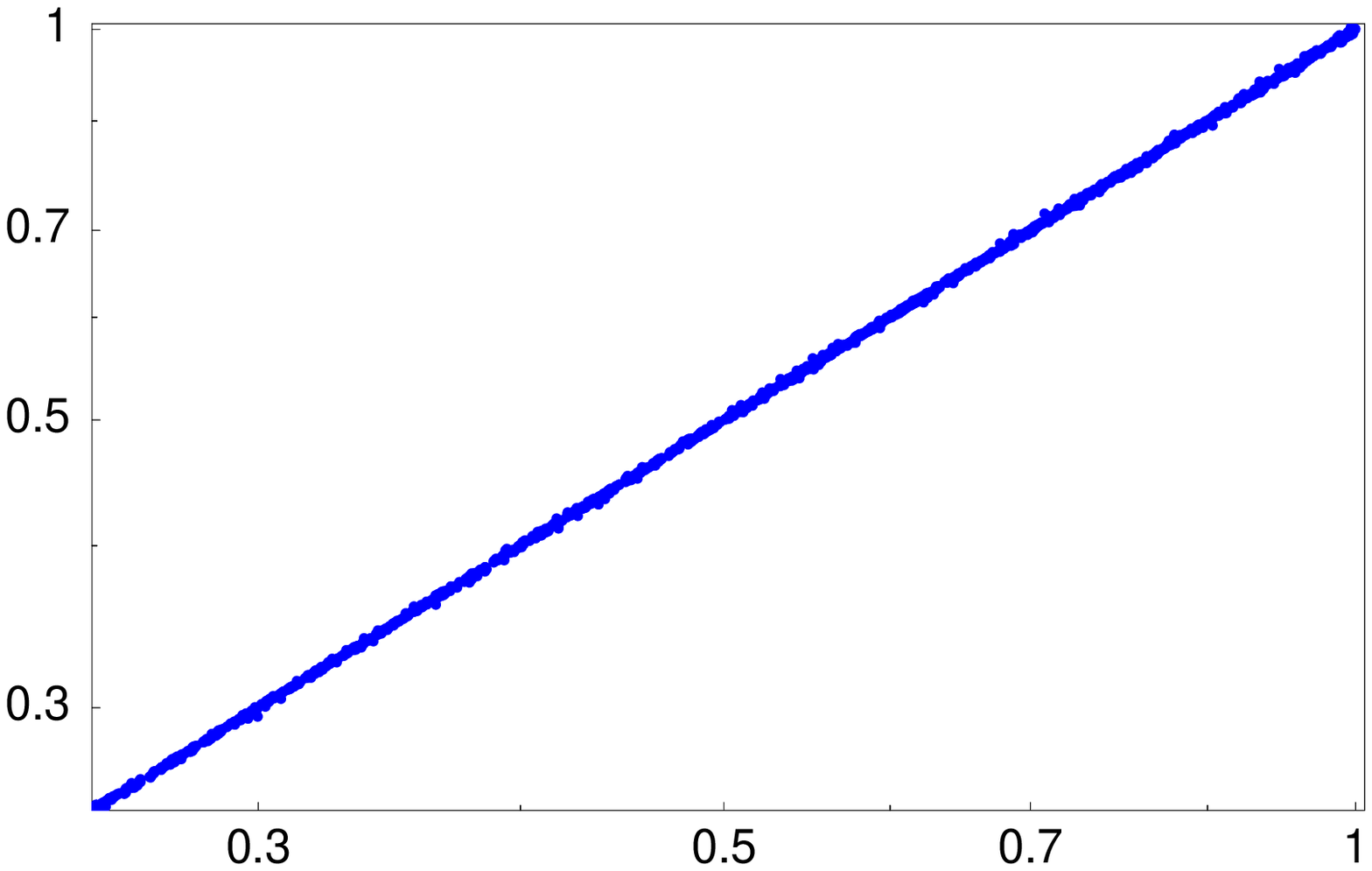}

\vskip-45mm
\begin{rotate}{90}
$Br({\tilde \tau}_1 \rightarrow e \sum \nu_i)/
Br({\tilde \tau}_1 \rightarrow \mu \sum \nu_i)$
\end{rotate}

\vskip5mm
\hskip90mm
$(\epsilon_1/\epsilon_2)^2$

\vskip-30mm
\hskip5mm
\epsfysize=140mm
\epsfxsize=100mm
\epsfbox{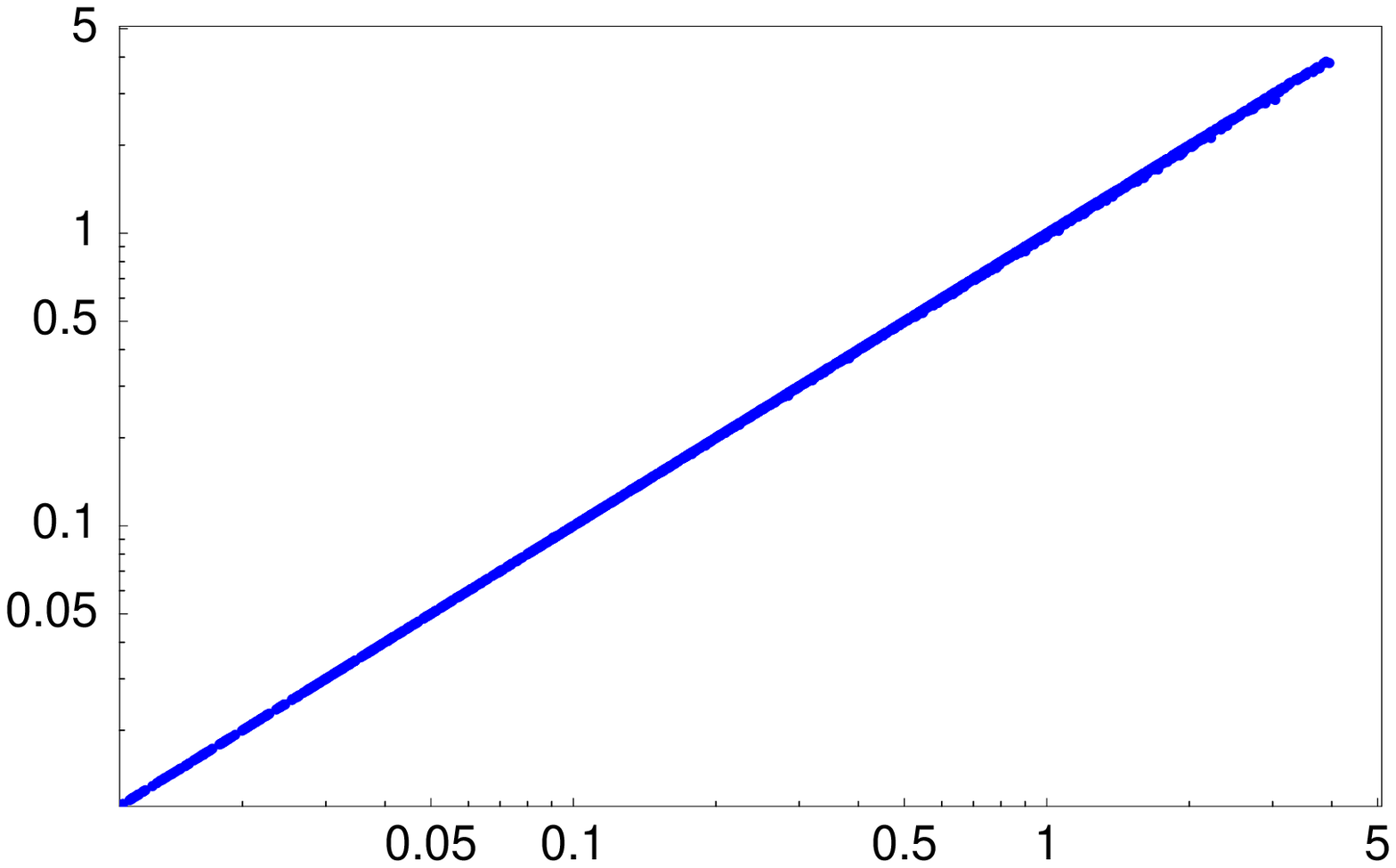}

\vskip-45mm
\begin{rotate}{90}
$Br({\tilde \mu}_1 \rightarrow e \sum \nu_i)/
Br({\tilde \mu}_1 \rightarrow \tau \sum \nu_i)$
\end{rotate}

\vskip5mm
\hskip90mm
$(\epsilon_1/\epsilon_3)^2$

\caption[]{Ratios of branching ratios for scalar tau (top panel) decays 
  versus $(\epsilon_1/\epsilon_2)^2$ and scalar muon decays (bottom
  panel) versus $(\epsilon_1/\epsilon_3)^2$ for Set3.}
\label{BrSet4b}
\end{figure}

As mentioned in the previous section, one expects that ratios of
branching ratios of various charged slepton decays contain rather
precise information on ratios of the bilinear parameters $\epsilon_i$.
That this is indeed the case is shown in \fig{BrSet4a} for the data of
Set2 and in \fig{BreSet4b} and \fig{BrSet4b} for the data of Set3.

As can be seen from these figures, the ratio of charged slepton
branching ratios are correlated with the ratios of the corresponding
BRpV parameters $\epsilon_i$, following very closely the expectation
from Eq. \ref{eq:SimSlLNu}, nearly insensitive to variation of the
other parameters.  Recall, that all the points were generated through
a rather generous scan over the mSUGRA parameters.  Ratios of
$\epsilon_i$'s should therefore be very precisely measurable.
Moreover, since only two of the three ratios of $\epsilon_i$'s are
independent it is possible to derive the following
\texttt{prediction}:

\smallskip
\centerline{$Br({\tilde \tau}_1 \rightarrow e \sum \nu_i)/
Br({\tilde \tau}_1 \rightarrow \mu \sum \nu_i)$ : 
$Br({\tilde \mu}_1 \rightarrow e \sum \nu_i)/
Br({\tilde \mu}_1 \rightarrow \tau \sum \nu_i)$ $\simeq$}
 
\centerline{$\simeq$ $Br({\tilde e}_1 \rightarrow \mu \sum \nu_i)/
Br({\tilde e}_1 \rightarrow \tau \sum \nu_i)$ }
\smallskip

\noindent
which provides an important cross check of the validity of our
bilinear R-parity model. Any significant departure from this equality
would be a clear sign that the bilinear model is incomplete.

\begin{figure}
\vskip-30mm
\hskip5mm
\epsfysize=140mm
\epsfxsize=100mm
\epsfbox{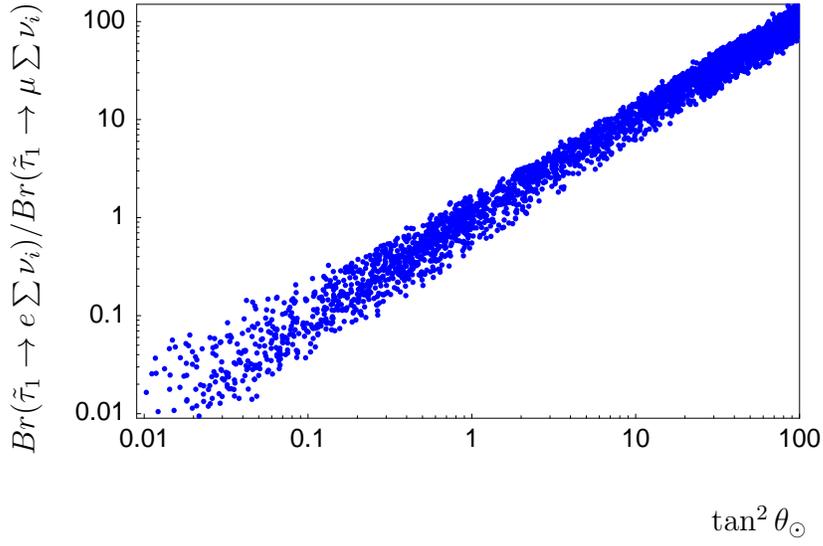}

\vskip-45mm
\begin{rotate}{90}
$Br({\tilde \tau}_1 \rightarrow e \sum \nu_i)/
Br({\tilde \tau}_1 \rightarrow \mu \sum \nu_i)$
\end{rotate}

\vskip5mm
\hskip90mm
$\tan^2\theta_{\odot}$
\vskip-30mm
\hskip5mm
\epsfysize=140mm
\epsfxsize=100mm
\epsfbox{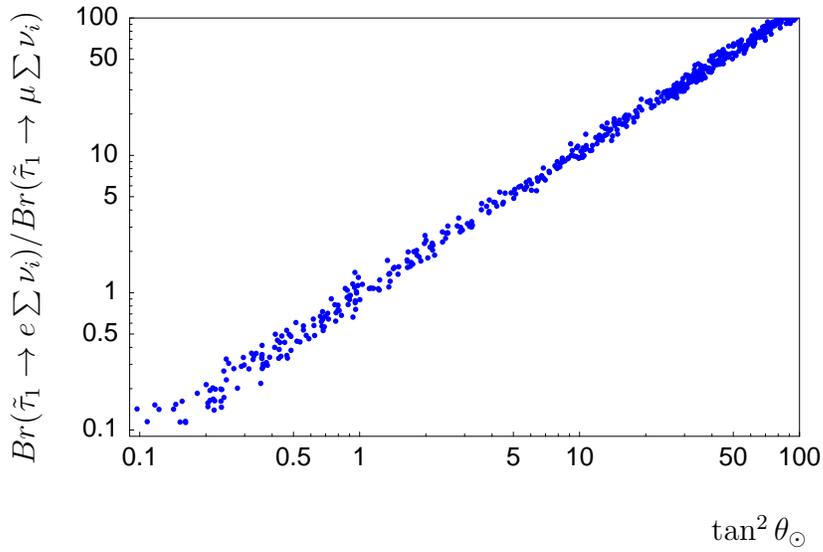}

\vskip-45mm
\begin{rotate}{90}
$Br({\tilde \tau}_1 \rightarrow e \sum \nu_i)/
Br({\tilde \tau}_1 \rightarrow \mu \sum \nu_i)$
\end{rotate}

\vskip5mm
\hskip90mm
$\tan^2\theta_{\odot}$

\caption[]{Ratios of branching ratios for scalar tau decays 
  versus $\tan^2\theta_{\odot}$ for Set2. The top panel shown all data
  points, the bottom one refers only to data points with
  $\epsilon_2/\epsilon_3$ restricted to the range [0.9,1.1].}
\label{BrSolar}
\end{figure}

As mentioned in the introduction current 
solar neutrino data prefer a large angle solution (LMA). In the BRpV
model the solar angle is mainly determined by the ratio
$\epsilon_1/\epsilon_2$ \cite{NuMass}.  A measured solar angle
therefore leads to a prediction for $Br({\tilde \tau}_1 \rightarrow e
\sum \nu_i)/ Br({\tilde \tau}_1 \rightarrow \mu \sum \nu_i)$, as shown
in \fig{BrSolar} for the data of Set2. With the current limits on
$\tan^2\theta_{\odot}$, which are $0.25 < \tan^2\theta_{\odot} < 0.83$
for the preferred LMA-MSW solution to the solar neutrino problem
\cite{Maltoni:2002ni} at 3 $\sigma$ \CL, one can currently predict
that this ratio in the BRpV model must be in the range $[0.09,1.8]$.
Additional input on $\epsilon_2/\epsilon_3$, for example
$\epsilon_2/\epsilon_3\simeq 1$ to within 10 \% would sharpen the
predicted value to $[0.15,1.1]$.  Obviously, also a more precise
measurement of the solar angle will lead to a tighter prediction in
the future. In this context it is worth noting that KamLAND
\cite{KamLand} should be able to fix the solar angle to within $\sim$
30 \%, if LMA is indeed the correct solution to the solar neutrino
problem.

Up to now we have discussed only \texttt{ratios} of R-parity violating
parameters, but charged scalar lepton decays allow, in principle, also
to gain information on \texttt{absolute values} of these parameters,
as relevant, e.~g. to fix the scale of neutrino masses determined
through the analysis of current solar and atmospheric
data~\cite{Maltoni:2002ni}.  However, such a measurement would require
at least some information on MSSM parameters which is at the moment 
unavailable.

\begin{figure}
\vskip-30mm
\hskip5mm
\epsfysize=140mm
\epsfxsize=100mm
\epsfbox{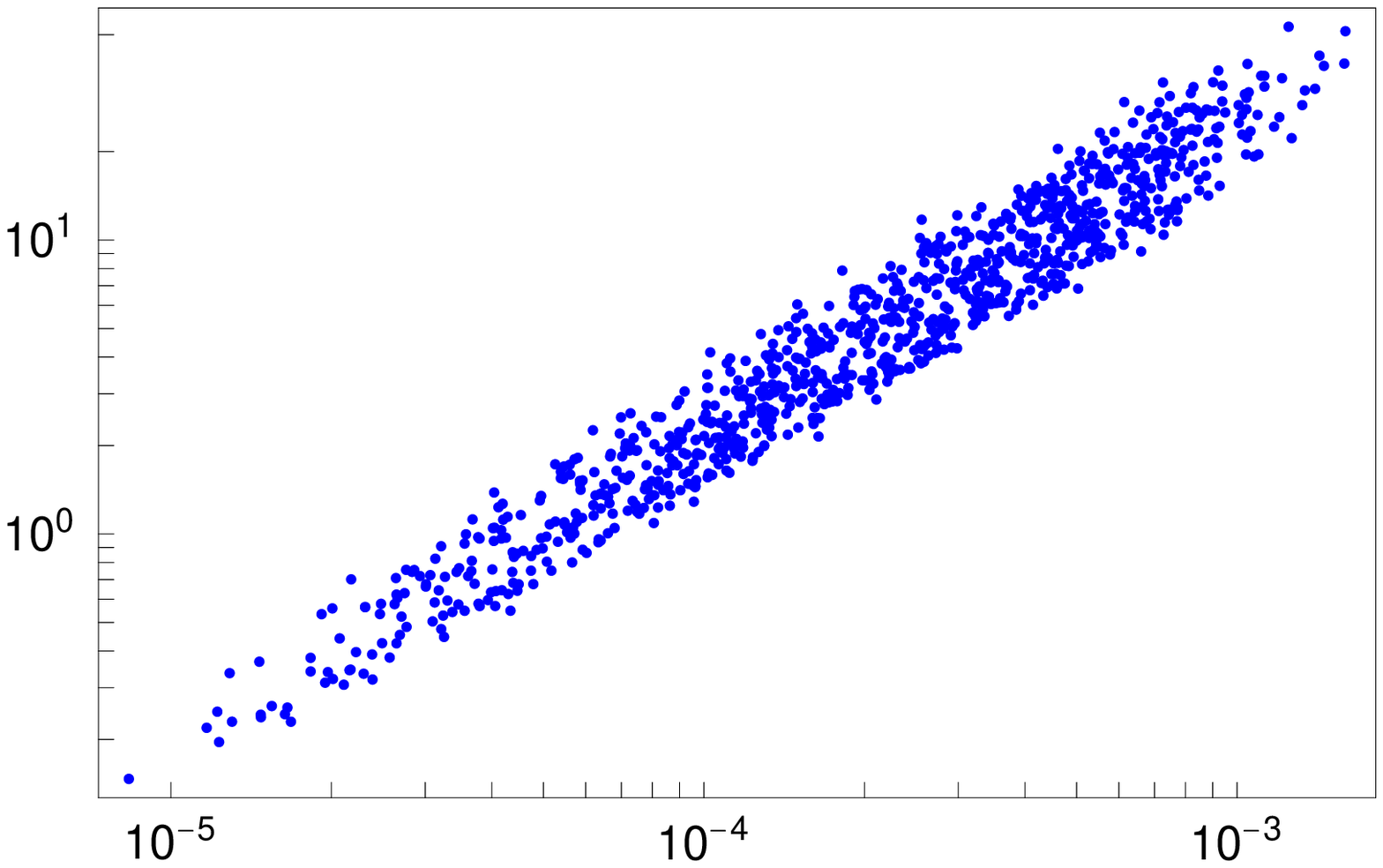}

\vskip-88mm
\begin{rotate}{90}
$\Gamma({\tilde \tau_1})$ [eV]
\end{rotate}

\vskip48mm
\hskip73mm
$(|\epsilon|/\mu)^2 m_{\tilde \tau_1}$ [GeV]

\vskip-30mm
\hskip5mm
\epsfysize=140mm
\epsfxsize=100mm
\epsfbox{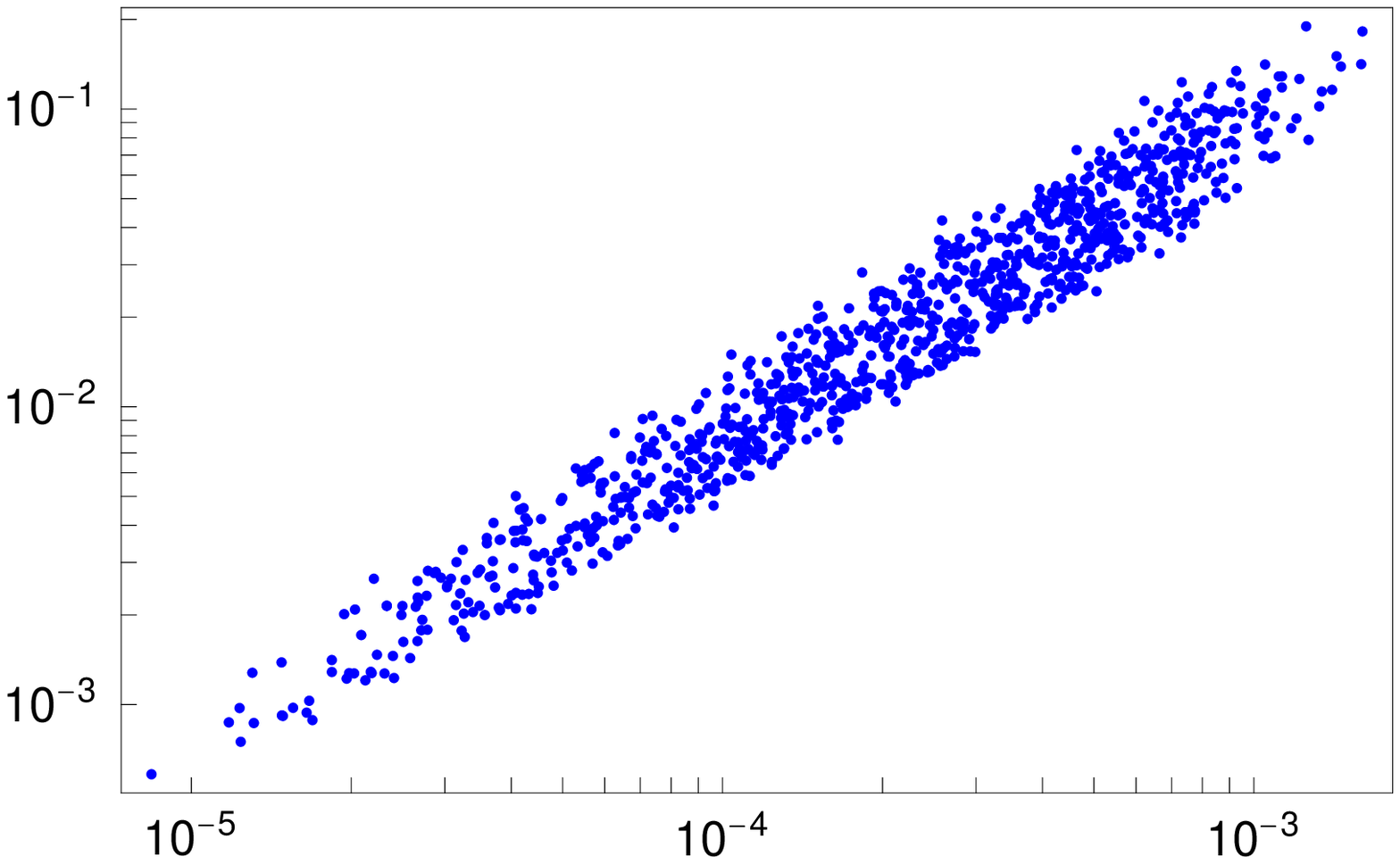}

\vskip-88mm
\begin{rotate}{90}
$\Gamma({\tilde \mu_1})$ [eV]
\end{rotate}

\vskip48mm
\hskip73mm
$(|\epsilon|/\mu)^2 m_{\tilde \mu_1}$ [GeV]

\caption[]{Total widths in [eV] for scalar tau decays (top) 
and scalar muon decays (bottom) for the data of Set1 versus 
$(|\epsilon|/\mu)^2 m_{\tilde l}$. Once $\mu$ and $m_{\tilde l}$ 
are measured, the widths provide information on the absolute 
value of $|\epsilon| \equiv |{\vec \epsilon}|$. Note that Set1 
fixes $\epsilon_2/\epsilon_3 \simeq 1$. In general, this ratio 
must be known with some accuracy, before a value for $|\epsilon|$ 
can be derived from the widths.}
\label{TotWidth}
\end{figure}

In \fig{TotWidth} we show the total widths in [eV] for scalar tau
decays (top panel) and scalar muon decays (bottom panel) for the data
of Set1 displayed versus $(|\epsilon|/\mu)^2 m_{\tilde l}$. Once $\mu$
and $m_{\tilde l}$ have been measured with some accuracy, one can
determine the absolute value of $|\epsilon|$ from this measurement,
provided $\epsilon_2/\epsilon_3$ is known (for example, from the ratio
$Br({\tilde e}_1 \rightarrow \mu \sum \nu_i)/ Br({\tilde e}_1
\rightarrow \tau \sum \nu_i)$).

\begin{figure}
\vskip-30mm
\hskip5mm
\epsfysize=140mm
\epsfxsize=100mm
\epsfbox{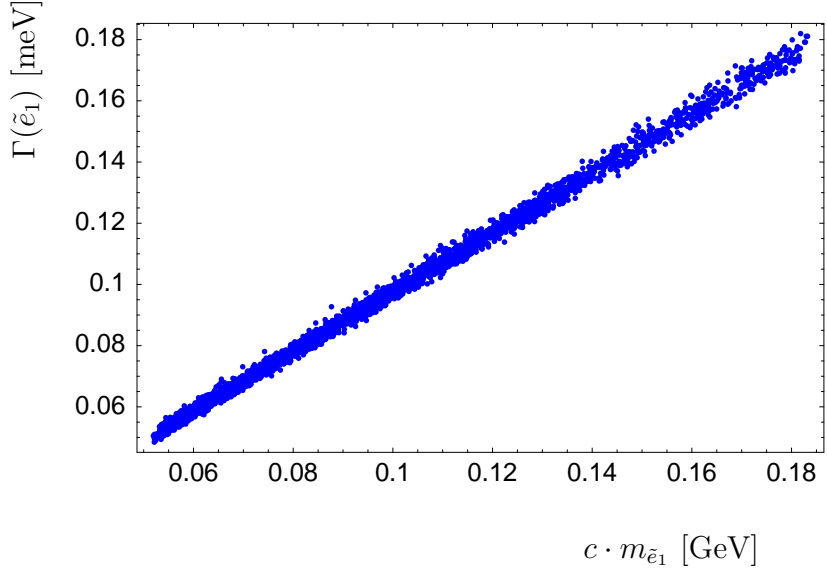}

\vskip-88mm
\begin{rotate}{90}
$\Gamma({\tilde e_1})$ [meV]
\end{rotate}

\vskip48mm
\hskip73mm
$c \cdot m_{\tilde e_1}$ [GeV]

\vskip-30mm
\hskip5mm
\epsfysize=140mm
\epsfxsize=100mm
\epsfbox{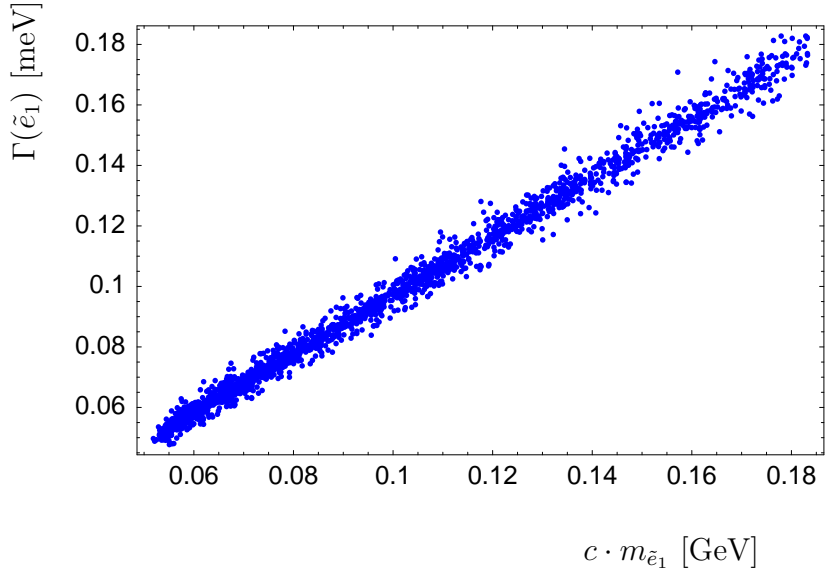}

\vskip-88mm
\begin{rotate}{90}
$\Gamma({\tilde e_1})$ [meV]
\end{rotate}

\vskip48mm
\hskip73mm
$c \cdot m_{\tilde e_1}$ [GeV]

\caption[]{Charged scalar electron total decay widths in [meV], the 
  top panel refers to Set2 while the bottom one is for Set3. The plots
  are versus $c \cdot m_{\tilde c_1}$, where $c=\frac{1}{8\pi} (g')^2
  x_1 |\Lambda|^2$.}
\label{SeWidth}
\end{figure}

In a similar way, the decay width of the scalar electron contains
information on $|{\vec \Lambda}|$, as is demonstrated in \fig{SeWidth}. A
priori knowledge on $\epsilon_2/\epsilon_3$ leads to a tighter
correlation, as can be seen from the comparison of the results for
Set2 and Set3.

To deduce the value of $|{\vec \Lambda}|$ from this measurement one
needs the parameter combination $x_1$, as defined in Eq.
(\ref{defx1}).  It contains the MSSM parameters $M_1$, $M_2$, $\mu$
and $\tan\beta$, which could be determined, for example, if at least
some of the neutralino and chargino eigenstates are accesible at the
LHC or a possible linear collider.

\section{Conclusions}
\label{sec:conclusions}

Supersymmetric models with bilinear R-parity breaking provide a
simple, testable framework for neutrino masses and mixings in
agreement with current solar, atmospheric and reactor neutrino
oscillation data. The model is testable at future colliders if the
neutralino is the LSP, as was shown previously, as well as in the
alternative case where one of the charged scalar leptons is the LSP,
as we have demonstrated here.

The measured neutrino mixing angles fix certain ratios of the bilinear
R-parity breaking parameters and, therefore, lead to well-defined
predictions for the ratio of branching ratios of certain slepton decay
modes, which should be easily measurable at a future collider such as
a high energy linear collider. Our main result is shown in
\fig{BrSolar}, where we display $Br({\tilde \tau}_1 \to e \sum \nu_i)/
Br({\tilde \tau}_1 \rightarrow \mu \sum \nu_i)$ versus the solar
neutrino angle, $\tan^2\theta_{\odot}$.

We have also shown how charged scalar lepton decays allow the
determination of other parameters of our model, thus providing a
definite test that bilinear R-parity breaking SUSY is the origin of
neutrino masses.

\section*{Acknowledgments}
  This work was supported by Spanish grants PB98-0693 and by the
  European Commission RTN network HPRN-CT-2000-00148.  M. H.  is 
  supported by a Spanish MCyT Ramon y Cajal contract. W.~P.~is supported 
by the 'Erwin Schr\"odinger fellowship No.~J2095' of the `Fonds zur
F\"orderung der wissenschaftlichen Forschung' of Austria FWF and
partly by the Swiss `Nationalfonds'.

\end{document}